\documentclass[twocolumn,showpacs,preprintnumbers,amsmath,amssymb,aps]{revtex4-1}
\usepackage{graphicx}
\usepackage{dcolumn}
\usepackage{bm}
\usepackage{float}
\usepackage{soul}

\begin{document}

\title{Discriminating two nonorthogonal states against a noise channel by feed-forward control}

\author{Li-Sha Guo$^{1}$}
\author{Bao-Ming Xu$^{1}$}
\author{Jian Zou$^{1}$}%
\email{zoujian@bit.edu.cn}
\author{Chao-Quan Wang$^{1}$}
\author{Hai-Li$^{2}$}
\author{Jun-Gang Li$^{1}$}%
\author{Bin Shao$^{1}$}%

\affiliation{$^{1}$School of Physics, Beijing Institute of Technology, Beijing 100081, China}%
\affiliation{$^{2}$School of Information and Electronic Engineering, Shandong Institute of Business and Technology, Yantai 264000, China}

\date{Submitted \today}

\begin{abstract}
We propose a scheme by using the feed-forward control (FFC) to realize a better effect of discrimination
of two nonorthogonal states after passing a noise channel based on the minimum-
error (ME) discrimination. We show that the application of
our scheme can highly improve the effect of discrimination compared with the ME discrimination without the FFC for any pair of nonorthogonal states and any degree of
amplitude damping (AD). Especially, the effect of our optimal discrimination can reach that of the two initial nonorthogonal pure states in the presence of the noise channel in a deterministic way
for equal a prior probabilities or even be better than that in a probabilistically way for unequal a
prior probabilities.

\end{abstract}


\maketitle

\section{Introduction}
In quantum information processing and quantum computing protocols, information is encoded in the state of a quantum system. After processing the information, it has to be read out or, in other words, the state of the system has to be determined. When the set of the possible output states is known and the states in the set are mutually orthogonal, they can be perfectly discriminated. However, when the possible output states are not orthogonal, they can not be always discriminated perfectly. With the rapid development of quantum information, the application of nonorthogonal states shows its widely significance in many areas. In particular, nonorthogonal states can be used in secure quantum cryptographic protocols \cite{Bennett1984,Bennett1992a,Bennett1992b}, most notably in the quantum key distribution (QKD) schemes, which are based on the two-state procedure developed by Bennett \cite{Bennett1992b}. So how to discriminate among nonorthogonal quantum states in an optimum way has gained much attention both theoretically and experimentally.
So far, there are mainly two approaches to optimally discriminate two nonorthogonal states. The first approach is the minimum-error (ME) discrimination \cite{Helstrom1976,Holevo1973,Yuen1975,Bergou2010}, in which inconclusive outcome is not allowed, hence errors are unavoidable. This scheme has been introduced by Helstrom \cite{Helstrom1976} originally and a general expression of the minimum achievable error probability for distinguishing between two mixed quantum states has also been obtained. For more than two mixed quantum states, there exists some analytical solutions only for some special quantum states \cite{Barnett2001,Andersson2002,Chou2003,Herzog2002}. The second approach is the so-called unambiguous discrimination (UD)
\cite{Ivanovic1987,Dieks1988,Peres1988a,Jaeger1995,Peres1998b,Duan1998,Chefles1998,Sun2002,Eldar2003b,Qiu2002}, first suggested by Ivanovic, Dicks, and Peres \cite{Ivanovic1987,Dieks1988,Peres1988a}, for the discrimination of two pure states. Unlike the ME discrimination, the UD allows an inconclusive result, but without any error and it just works for linearly independent states. So far, the UD has mostly focused on pure states. For mixed states, only some useful bounds on the total success probability, together with some useful reduction theorems, have been presented \cite{Sun2002,Rudolph2003,Feng2004,Raynal2003,Herzog2005,Raynal2005,Bergou2006}.

As mentioned above, quantum communication is highly dependent on two nonorthogonal states, such as the well-known B92 quantum key distribution protocol \cite{Bennett1992b}, which is based on the discrimination of two nonorthogonal states. And all the existing quantum cryptographic schemes are usually based on pure states. However, due to the unavoidable coupling with the environment, decoherence, as a detrimental factor in quantum communication, can not be completely removed, which would turn the initial nonorthogonal pure states into mixed states after their transmission.

In this paper, we consider a realistic situation: Alice first encodes the information in two nonorthogonal pure states, and then sends them to Bob through a noise channel. Because of the noise channel, the nonorthogonal states sent by Alice would become mixed, so Bob has to discriminate between two nonorthogonal mixed states. The above consideration has its practical significance, because the commercial fibers are widely used in quantum communication, which would introduce decoherence and affect the purities of the transmitted states more or less inevitably. What we concern in this paper is how to improve the effect of discrimination of two
nonorthogonal states after passing an amplitude damping noise (AD) channel. To this end, we propose a scheme by using the feed-forward control (FFC) \cite{Wang2014} to realize a better effect of discrimination. Note that the FFC in our scheme is based on an instantaneous measurement, while some other
relevant works of quantum-state discrimination in recent papers \cite{Jacobs2007,Muller2012,Becerra2011} are making use of the feed forward or feedback control based on continuous measurement. Besides, the purpose of the usage of feed forward or feedback control is different between our work and their works. For the present paper, it managed to utilize the FFC to overcome the environmental noise in the state transmission process, while the other three papers are not for this purpose.
Specifically, in Ref. \cite{Jacobs2007}, Jacobs considered a continuous implementation of the optimal measurement for distinguishing between two nonorthogonal states, and showed that the feedback control used during this measurement can increase the transformation rate of the channel. While Refs. \cite{Muller2012,Becerra2011} are making use of successive measurements on parts of the state and feed forward to achieve an error rate below the heterodyne limit for multiple state discrimination.

\begin{figure}[tbp]
\begin{center}
\includegraphics[width=8cm]{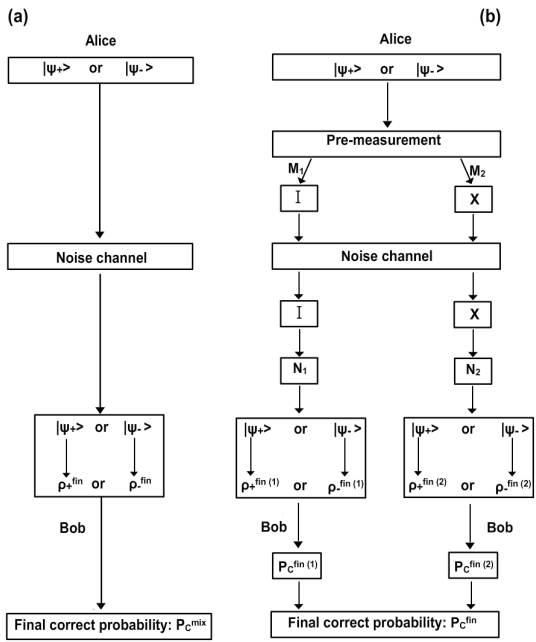}
\caption{A schematic diagram showing the procedure of discrimination between two nonorthogonal mixed states in quantum communication. (a) The ME discrimination without the FFC (conventional scheme). (b) The ME discrimination based on the FFC (our scheme).}
\end{center}
\end{figure}

The procedure of our scheme is like this: Alice prepares a qubit in one of two known pure states, $|\psi_{+}\rangle$ or $|\psi_{-}\rangle$ with a prior probabilities $q_{+}$ and $q_{-}$ respectively. Before sending the signal to Bob, Alice should use our pre-weak measurement and unitary operation to prepare the state into the one almost immune to the noise channel, and then sends it to Bob through the noise channel and tells Bob the pre-weak measurement result through classical communication. When Bob receives the signal, he can use our reversed unitary operation and the reversed post-weak measurement to restore the information of the pure state sent by Alice originally, and then further determine what the quantum state sent is without inconclusive results, i.e., Bob can perform the ME discrimination on the two mixed states. Our scheme is universal for any pair of nonorthogonal states and any degree of amplitude damping, i.e., by using our scheme, the effect of discrimination of two nonorthogonal states after passing the noise channel can be highly improved compared with the ME discrimination without the FFC for any pair of initial nonorthogonal states. And the smaller the overlap of the two nonorthogonal states (the heavier the amplitude damping) is, the more advantage our scheme has. In particular, the effect of our optimal discrimination can be equivalent to (even better than) that of the two initial nonorthogonal pure states in a deterministic way when $q_{+}=q_{-}=1/2$ (in a probabilistically way when $q_{+}\neq q_{-}$), that is to say, the optimal discrimination can completely counteract the impact of the noise channel on the effect of discrimination of two nonorthogonal states or even improve the effect of discrimination of two initial nonorthogonal pure states.

This paper is structured as follows. In the next section, we introduce our scheme by using the FFC to realize a better effect of discrimination of two nonorthogonal states after passing the AD channel, and based on which we consider two cases in the following. Specifically, in Sec. \uppercase\expandafter{\romannumeral3}, we consider the first case of discrimination based on state protection and compare our result with that of the ME discrimination without the FFC. In Sec. \uppercase\expandafter{\romannumeral4}, we consider the second case of optimal discrimination and also compare our result with that of the ME discrimination without the FFC. Finally, in Sec. \uppercase\expandafter{\romannumeral5} we give the conclusion of our results.
\section{Scheme based on the FFC to improve the effect of discrimination}
In this section, we begin with reviewing the ME discrimination \cite{Helstrom1976,Holevo1973,Yuen1975,Bergou2010} which requires conclusive discrimination all the time, thus errors are unavoidable.
When we minimize (or maximize) the probability with which we make a wrong guess (or a correct guess) of what the quantum state actually is,
it will lead to the so-called ME discrimination, and the corresponding probability is called the minimum error probability $P_{E}$ (or the maximum correct probability $P_{C}$). In this paper, we only consider the two-state discrimination. Assume that a quantum system is in either the state $\rho_{+}$ or $\rho_{-}$
with the prior probabilities $q_{+}$ and $q_{-}$
respectively, thus it can be described as:
\begin{equation}\label{rhoo}
    \rho=q_{+}\rho_{+}+q_{-}\rho_{-}.
\end{equation}
The minimum error probability, $P_{E}$, or the maximum correct probability, $P_{C}=1-P_{E}$ for discriminating between $\rho_{+}$ and $\rho_{-}$ with the corresponding prior probabilities $q_{+}$ and $q_{-}$ was
derived by Helstrom \cite{Helstrom1976} in the framework of quantum detection and estimation theory, i.e., the so called Helstrom limit
\begin{equation}\label{PCc}
      P_{C}=\frac{1}{2}(1+\sum_{k}|\lambda_{k}|),
\end{equation}
which can be realized by means of the optimal measurement,
where $\lambda_{k}$ are the eigenvalues of the operator $ \Lambda=q_{-}\rho_{-}-q_{+}\rho_{+}$.
In the special case that the states to be distinguished
are pure states $|\psi_{+}\rangle$ and $|\psi_{-}\rangle$, this expression
reduces to
\begin{equation}\label{ppcure}
    P_{C}=\frac{1}{2}(1+\sqrt{1-4q_{+}q_{-}|\langle\psi_{+}|\psi_{-}\rangle|^{2}}).
\end{equation}

In the following we would apply this ME discrimination in the context of a communication scenario of two parties, Alice and Bob. First, an ensemble of quantum systems are prepared so that each individual system is
prepared in one of two known states, $|\psi_{+}\rangle$ or $|\psi_{-}\rangle$ with
the prior probability $q_{+}$ or $q_{-}$, respectively. So the quantum system can be described as:
\begin{equation}\label{rhoin}
  \rho^{in}=q_{+}|\psi_{+}\rangle\langle \psi_{+}|+q_{-}|\psi_{-}\rangle\langle \psi_{-}|,
\end{equation}
where
\begin{equation}\label{psi}
|\psi_{\pm}\rangle=\cos\frac{\theta}{2}|+\rangle\pm \sin\frac{\theta}{2}|-\rangle,
\end{equation}
with $|\pm\rangle=(|0\rangle\pm|1\rangle)/\sqrt{2}$ and $\theta\in[0,\pi/2]$.
And then Alice would hand it over to Bob, whose task is to determine which one of the two states he is given. He also knows how the ensemble were prepared, i.e., he has full knowledge of the two possible states and their prior probabilities but does not know the actual state that was drawn. All he can do is to perform a single measurement or perhaps a POVM on the individual system he receives.

According to Eq. (\ref{ppcure}),
the maximum correct probability for discriminating between the above two pure states is
\begin{equation}\label{pcpure}
    P_{C}^{pure}=\frac{1}{2}(1+\sqrt{1-4q_{+}q_{-}\cos^2\theta}).
\end{equation}
If we assume that $q_{+}$=$q_{-}$=1/2, then $P_{C}^{pure}$ can be simplified as:
\begin{equation}\label{pcpure=}
    P_{C}^{pure}=\frac{1}{2}(1+\sin\theta).
\end{equation}

However, the noise caused by environment in communication can not be ignored so we have to deal with two nonorthogonal mixed states. That is, in the presence of noise, we have to discriminate between the two transmitted mixed states $\rho_{+}^{fin}$ and $\rho_{-}^{fin}$ to attain the original goal of discriminating between $|\psi_{+}\rangle$ and $|\psi_{-}\rangle$ (see Fig. 1(a)).
In this paper, we only consider the AD channel, and for this channel the evolution of any input state $\rho^{in}$ can be represented by the sum of Kraus operators:
\begin{equation}\label{noise}
  \rho^{fin}=\varepsilon(\rho^{in})=\sum_{k=1,2}E_{k}\rho^{in}E_{k}^{\dagger},
\end{equation}
where
\begin{equation}
E_{1}=\left( \begin{smallmatrix}
1& 0\\ 0&\sqrt{1-r}
\end{smallmatrix} \right),~
E_{2}=\left( \begin{smallmatrix}
0& \sqrt{r}\\ 0&0
\end{smallmatrix} \right)
\end{equation}
in the computational basis $\{|0\rangle,|1\rangle\}$, and $r$ represents the magnitude of decoherence.
Substituting Eq. (\ref{rhoin}) into Eq. (\ref{noise}) we can obtain the final state after passing the AD channel:
\begin{equation}\label{rhofin}
  \rho^{fin}=q_{+}\rho_{+}^{fin}+q_{-}\rho_{-}^{fin},
\end{equation}
where
\begin{equation}\label{rho}
    \rho_{\pm}^{fin}=
\begin{pmatrix}
1-\frac{1}{2}(1\mp\sin\theta)(1-r)& \frac{1}{2}\cos\theta\sqrt{1-r}  \\
\\
\frac{1}{2}\cos\theta\sqrt{1-r} & \frac{1}{2}(1\mp\sin\theta)(1-r)
\end{pmatrix}.
\end{equation}

Our aim is to discriminate between $\rho_{+}^{fin}$ and $\rho_{-}^{fin}$£¬which correspond to the input states $|\psi_{+}\rangle$ and $|\psi_{-}\rangle$, with a prior probabilities $q_{+}$ and $q_{-}$ respectively.
According to Eq. (\ref{PCc}), we can obtain the maximum correct probability under the conventional scheme (see Fig. 1(a)):
\begin{equation}\label{PCCC}
     P_{C}^{mix}=\frac{1}{2}(1+\frac{1}{2}|\chi-\xi|+\frac{1}{2}|\chi+\xi|),
\end{equation}
where $\xi=\sqrt{(1-r)(\chi\cos\theta)^{2}+((1-r)\sin\theta+\chi r)^{2}}$ with $\chi=q_{+}-q_{-}$.
If we assume that $q_{+}$=$q_{-}$=1/2, then the maximum correct probability for discriminating between $\rho_{+}^{fin}$ and $\rho_{-}^{fin}$ can be simplified as:
\begin{equation}\label{PC}
\begin{split}
    P_{C}^{mix} =&\frac{1}{2}(1+\sin\theta)-\frac{1}{2}r\sin\theta\\
    =&P_{C}^{pure}-\frac{1}{2}r\sin\theta.
\end{split}
\end{equation}
Because $(1/2)r\sin\theta>0$, it is obvious that $P_{C}^{mix}$ is smaller than $P_{C}^{pure}$ (Eq. (\ref{pcpure=})) except for $r=0$. That is to say, noise in this channel generally reduces the effect (maximum correct probability) of discrimination.

In what follows, we introduce our scheme by using the FFC to improve the effect (maximum correct probability) of discrimination of two nonorthogonal states after passing the AD channel (see Fig. 1(b)). Generally speaking, the whole procedure for our scheme can be divided into four steps:

1) The input state Alice would send to Bob is $ \rho^{in}=q_{+}|\psi_{+}\rangle\langle \psi_{+}|+q_{-}|\psi_{-}\rangle\langle \psi_{-}|$
as discussed above. Before sending it to Bob, first Alice would perform our complete pre-weak measurement $\{M_{1},M_{2}\}$, which are represented respectively as
\begin{equation}
M_{1}=\left( \begin{smallmatrix}
\sqrt{p}& 0\\ 0&\sqrt{1-p}
\end{smallmatrix} \right),~
M_{2}=\left( \begin{smallmatrix}
\sqrt{1-p}& 0\\ 0&\sqrt{p}
\end{smallmatrix} \right),
\end{equation}
with $\sum_{i=1}^{2}M_{i}^{\dagger} M_{i}=I$, and $p\in[0,1]$ is the pre-weak measurement strength. Note that this complete pre-weak measurement can be
realized theoretically by coupling the system to a meter and performing
the usual projective measurements on the meter only \cite{Audretsch2001}.
Experimentally, the weak measurement on the signal qubit (photon) has been realized by performing a full-strength projective measurement on the meter qubit (photon), which is entangled with the signal qubit. The strength (ranging from 0 to 1) of the weak measurement on the signal qubit can be determined by the input meter state \cite{Gillett2010}.

If result $i$ ($i=1,2$) is acquired,
the state can be expressed as
\begin{equation}\label{rhomi}
  \rho_{i}=\frac{M_{i}\rho^{in}M_{i}^{\dagger}}{\mathrm{Tr}[M_{i}^{\dagger}M_{i}\rho^{in}]},
\end{equation}
where
\begin{equation}\label{pmi}
  P_{M_{i}}=\mathrm{Tr}[M_{i}^{\dagger}M_{i}\rho^{in}]=\sum_{n=+,-}q_{n}\langle\psi_{n}|M_{i}^{\dagger}M_{i}|\psi_{n}\rangle
\end{equation}
is the corresponding success (selective) probability.
The state (\ref{rhomi}) can be divided into two parts with respect to $|\psi_{+}\rangle$ and $|\psi_{-}\rangle$:
\begin{equation}\label{rhoi}
  \rho_{i}=q_{i+}\rho_{i+}+q_{i-}\rho_{i-},
\end{equation}
where
\begin{equation}\label{rhoipm}
  \rho_{i\pm}=\frac{M_{i}|\psi_{\pm}\rangle\langle\psi_{\pm}|M_{i}^{\dagger}}{\langle\psi_{\pm}|M_{i}^{\dagger}M_{i}|\psi_{\pm}\rangle}
\end{equation}
is the state after the measurement $M_{i}$ acting on $|\psi_{\pm}\rangle\langle\psi_{\pm}|$ and
\begin{equation}\label{qipm}
  q_{i\pm}=\frac{q_{\pm}\langle\psi_{\pm}|M_{i}^{\dagger}M_{i}|\psi_{\pm}\rangle}{P_{M_{i}}}
\end{equation}
is the modified prior probability (by $M_{i}$) corresponding to state $\rho_{i\pm}$.

And then according to the pre-weak measurement result $i$, Alice would perform the corresponding unitary operation $U_{i}$ with $U_{1}=I, ~U_{2}=\sigma_{x}$. More specifically, if result 1 is obtained Alice would do nothing on the qubit before it passes the noise channel; If result 2 is acquired, Alice would perform the $X$ gate on the qubit. After the unitary operation the state can be expressed as
 \begin{equation}\label{rho'i}
   \rho^{'}_{i}=q_{i+}\rho^{'}_{i+}+q_{i-}\rho^{'}_{i-},
\end{equation}
with $\rho^{'}_{i\pm}=U_{i}\rho_{i\pm}U^{\dag}_{i}$.

2) And then Alice would send the qubit to Bob through the noise channel and tell Bob the pre-weak measurement result $i$ through classical communication. After the noise channel the state can be expressed as
\begin{equation}\label{rho''i}
   \rho^{''}_{i}=q_{i+}\varepsilon(\rho^{'}_{i+})+q_{i-}\varepsilon(\rho^{'}_{i-}),
\end{equation}
where $\varepsilon(\cdot)=\sum_{k=1,2}E_{k}\cdot E_{k}^{\dagger}$ is the same as Eq. (\ref{noise}).

3) According to the pre-weak measurement result $i$ told by Alice, Bob would first perform the corresponding unitary operation $U_{i}$ on the state $\rho^{''}_{i}$: If result 1 occurs, Bob would do nothing on the qubit; If result 2 is acquired, Bob would perform the $X$ gate on the qubit. Then the state becomes
\begin{equation}\label{rho'''i}
\begin{split}
   \rho^{'''}_{i}&=q_{i+}U_{i}\varepsilon(\rho^{'}_{i+})U^{\dag}_{i}+q_{i-}U_{i}\varepsilon(\rho^{'}_{i-})U^{\dag}_{i} \\
                 &=q_{i+}\rho^{'''}_{i+}+q_{i-}\rho^{'''}_{i-}.
\end{split}
\end{equation}
Afterwards Bob would perform the reversed post-weak measurement $N_{i}$ on the qubit state $\rho^{'''}_{i}$ which are represented respectively as
\begin{equation}
N_{1}=\left( \begin{smallmatrix}
\sqrt{1-p_{1}}& 0\\ 0&1
\end{smallmatrix} \right),~
N_{2}=\left( \begin{smallmatrix}
1& 0\\ 0&\sqrt{1-p_{2}}
\end{smallmatrix} \right),
\end{equation}
where $p_{i}\in[0,1]$ ($i=1,2$) is the post-weak measurement strength.
Note that $N_{1}$ and $N_{2}$ are resulting from different sets of POVM measurement, i.e., $N_{1}$ belongs to the complete measurement set \{$N_{1}$, ~$\bar{N}_{1}$\} and $N_{2}$ belongs to the complete measurement set \{$N_{2}$, ~$\bar{N}_{2}$\}, i.e., $N_{1}^{\dagger} N_{1}+\bar{N}_{1}^{\dagger} \bar{N}_{1}=I$ and $N_{2}^{\dagger} N_{2}+\bar{N}_{2}^{\dagger} \bar{N}_{2}=I$. According to the pre-weak measurement result $i$, Bob just preserves result $N_{i}$ and discards result $\bar{N}_{i}$.
Moreover, the incomplete weak measurement has also been experimentally realized in Ref. \cite{Kim2012}. Specifically, the weak measurement is carried out on
the polarization qubit using Brewster-angle glass plates (BPs)
with a measurement strength which can be varied by changing the number of BPs.

Similar to Eqs. (\ref{rhomi})-(\ref{qipm}), the success (selective) probability after the post-weak measurement $N_{i}$ acting on the state $\rho^{'''}_{i}$ can be expressed as
\begin{equation}\label{pni}
  P_{N_{i}}=\mathrm{Tr}[N_{i}^{\dagger}N_{i}\rho^{'''}_{i}]
\end{equation}
and the selected state after the measurement $N_{i}$ can be expressed as
\begin{equation}\label{rhofini}
  \rho^{fin(i)}=q^{fin(i)}_{+}\rho^{fin(i)}_{+}+q^{fin(i)}_{-}\rho^{fin(i)}_{-}
\end{equation}
where
\begin{equation}\label{rhifinipm}
  \rho^{fin(i)}_{\pm}=\frac{N_{i}\rho^{'''}_{i\pm}N_{i}^{\dagger}}{\mathrm{Tr}[N_{i}^{\dagger}N_{i}\rho^{'''}_{i\pm}]}
\end{equation}
is the state after the measurement $N_{i}$ acting on $\rho^{'''}_{i\pm}$ and
\begin{equation}\label{qfinipm}
  q^{fin(i)}_{\pm}=\frac{q_{i\pm}\mathrm{Tr}[N_{i}^{\dagger}N_{i}\rho^{'''}_{i\pm}]}{P_{N_{i}}}
\end{equation}
is the modified prior probability (by $M_{i}$ and $N_{i}$) corresponding to state $\rho^{fin(i)}_{\pm}$.

Note that the success (selective) probability $P_{S_{i}}$ of obtaining the state $\rho^{fin(i)}$ (Eq. (\ref{rhofini})) resulting from the initial state $\rho^{in}$ (Eq. (\ref{rhoin})) is not equivalent to $P_{N_{i}}$ (Eq. (\ref{pni})). Actually $P_{S_{i}}$ is due to the twice selections, the first is caused by the pre-weak measurement $M_{i}$ with the selection probability $P_{M_{i}}$ (Eq. (\ref{pmi})) and the second is caused by the post-weak measurement $N_{i}$ with the selection probability $P_{N_{i}}$ (Eq. (\ref{pni})), so it should be expressed as:
\begin{equation}\label{ps1}
\begin{split}
  P_{S_{i}}&=P_{M_{i}}P_{N_{i}}\\
 & =q_{-}[\eta_{i}p(1-p_{i})+(1-\eta_{i})(1-p)(1-p_{i}r)]\\
&+q_{+}[(1-\eta_{i})p(1-p_{i})+\eta_{i}(1-p)(1-p_{i}r)]
\end{split}
\end{equation}
with $\eta_{1}=(1-\sin\theta)/2,~\eta_{2}=(1+\sin\theta)/2$.

4) After obtaining the state $\rho^{fin(i)}$ (Eq. (\ref{rhofini})) with a success (selective) probability $P_{S_{i}}$ (Eq. (\ref{ps1})), Bob would discriminate the states $\rho_{+}^{fin(i)}$ and $\rho_{-}^{fin(i)}$, using the ME discrimination, with the corresponding modified prior probabilities $q_{+}^{fin(i)}$ and $q_{-}^{fin(i)}$ according to the pre-weak measurement result $i$ ($i=1,2$). According to Eq. (\ref{PCc}), the maximum correct probability for discriminating between $\rho_{+}^{fin(i)}$ and $\rho_{-}^{fin(i)}$ ($i=1,2$) can be expressed as:
\begin{equation}\label{PC12}
\begin{split}
      P_{C}^{fin(i)}=\frac{1}{2}(1&+\frac{1}{2P_{S_{i}}}|g_{i}+\sqrt{4b_{i}^{2}+(g_{i}-2a_{i})^{2}}|\\
      &+\frac{1}{2P_{S_{i}}}|g_{i}-\sqrt{4b_{i}^{2}+(g_{i}-2a_{i})^{2}}|),
\end{split}
\end{equation}
where $a_{i}=q_{-}[\eta_{i}p(1-p_{i})+(1-\eta_{i})(1-p)(1-p_{i})r]-q_{+}[(1-\eta_{i})p(1-p_{i})+\eta_{i}(1-p)(1-p_{i})r];~b_{i}
=1/2(q_{-}-q_{+})\cos\theta\sqrt{p(1-p)(1-p_{i})(1-r)};~g_{i}=q_{-}[\eta_{i}p(1-p_{i})+(1-\eta_{i})(1-p)(1-p_{i}r)]
-q_{+}[(1-\eta_{i})p(1-p_{i})+\eta_{i}(1-p)(1-p_{i}r)]$.

At last, we define our final maximum correct probability of the whole discrimination process to be the weighted average of the two maximum correct probabilities $P_{C}^{fin(1)}$ and $P_{C}^{fin(2)}$, with the weights $P_{S_{1}}/(P_{S_{1}}+P_{S_{2}})$ and $P_{S_{2}}/(P_{S_{1}}+P_{S_{2}})$ respectively:
\begin{equation}\label{PCFIN}
    P_{C}^{fin}=\frac{P_{S_{1}}P_{C}^{fin(1)}+P_{S_{2}}P_{C}^{fin(2)}}{P_{S_{1}}+P_{S_{2}}},
\end{equation}
and define our final success probability of our scheme as:
\begin{equation}\label{PS}
    P_{S}= P_{S_{1}}+P_{S_{2}}.
\end{equation}
That is, we consider our scheme to be successful only when we obtain either the state $\rho^{fin(1)}$ or $\rho^{fin(2)}$ (Eq. (\ref{rhofini})) from the initial state $\rho^{in}$ (Eq. (\ref{rhoin})).

If we assume that $q_{+}$=$q_{-}$=$1/2$, which is widely considered in state discrimination field, then the above two expressions of $P_{C}^{fin}$ and $P_{S}$ can be simplified as

\begin{equation}\label{PCFIN=}
    P_{C}^{fin}=\frac{1}{2}+\frac{|p-(1-p)r|(2-p_{1}-p_{2})+2(1-p)(1-r)}{2\sin^{-1}\theta(2-(p_{1}+p_{2})(p(1-r)+r))},
\end{equation}
and
\begin{equation}\label{PS=}
   P_{S}=\frac{1}{2}(2-(p_{1}+p_{2})(p(1-r)+r)).
\end{equation}
From the above Eqs. (\ref{PCFIN=}), (\ref{PS=}), we can see that the success probability does not depend on the initial nonorthogonal states but the maximum correct probability is state dependent.
\section{Discrimination based on state protection }
Similar to the discussions in Refs. \cite{Wang2014,Korotkov2010}, by using the quantum trajectory theory, we obtain the following conditions: \begin{equation}\label{p1}
    p_{1}=1-\frac{(1-p)(1-r)}{p},
\end{equation}
and
\begin{equation}\label{p2}
    p_{2}=1-\frac{(1-p)(1-r)}{p},
\end{equation}
under which the final state along the ``no jumping" trajectory will be the initial state. Especially when the pre-measurement strength $p\rightarrow1$, $\rho_{\pm}^{fin(i)}\rightarrow|\psi_{\pm}\rangle\langle \psi_{\pm}|$ ($i=1,2$), hence the maximum correct probability for discriminating between the two nonorthogonal states after passing the noise channel is close to that of the two initial nonorthogonal pure states when $p\rightarrow1$. It should be noted that in this case, the post-weak measurements $N_{1}$ and $N_{2}$ are approximately the reversal of the corresponding pre-weak measurements $M_{1}$ and $M_{2}$, i.e., $N_{1}M_{1}\sim I$ and $N_{2}M_{2}\sim I$, as a result, the information about the initial state can be almost restored.
So we call it the discrimination based on state protection and define Eqs. (\ref{p1}), (\ref{p2}) as the state protection conditions. It can be seen from Eqs. (\ref{p1}), (\ref{p2}) that because $p_{1}$ and $p_{2}\geq0$, $p\in[\frac{1-r}{2-r},1]$.

First we consider the general case $q_{+}$=$q_{-}$=$1/2$ for the discrimination based on state protection. Substituting Eqs. (\ref{p1}), (\ref{p2}) into Eqs. (\ref{PCFIN=}), (\ref{PS=}), we can obtain the final maximum correct probability and the final success probability:
\begin{equation}\label{PCFIN=1}
     P_{C}^{fin}=\frac{1}{2}(1+\sin\theta\frac{p+|p-(1-p)r|}{2p+(1-p)r}),
\end{equation}
and
\begin{equation}\label{PS=1}
    P_{S}=(1-p)(1-r)(1+\frac{p+(1-p)r}{p}).
\end{equation}
\begin{center}
\includegraphics[width=8cm]{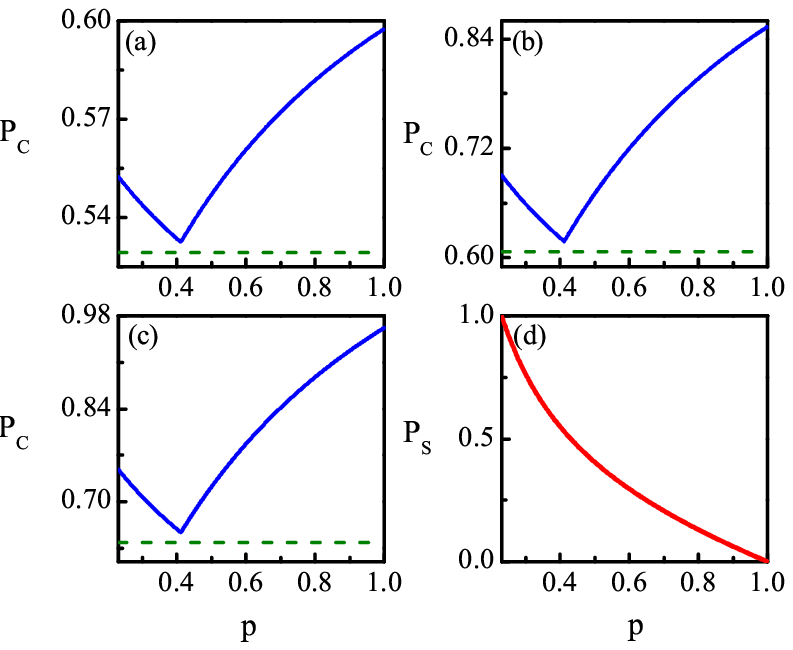}
\parbox{8cm}{\small{FIG. 2.} (Color online) The maximum correct probabilities $P_{C}^{fin}$ (Eq. (\ref{PCFIN=1})) (solid-blue lines) and $P_{C}^{mix}$ (Eq. (\ref{PC})) (dashed-olive lines) as functions of the measurement strength $p$ for (a) $\theta=\pi/16$; (b) $\theta=4\pi/16$ and (c) $\theta=6\pi/16$. Panel (d) plots the success probability $P_{S}$ corresponding to panels (a--c) for our scheme which is state independent. $r=0.7, ~q_{+}=q_{-}=1/2$.}
\end{center}

As an example, we first plot the maximum correct probability and the success probability for our scheme as functions of the measurement strength $p$ in Fig. 2 for different pairs of nonorthogonal states, with a fixed amplitude damping $r=0.7$ and equal a prior probabilities $q_{+}=q_{-}=1/2$. We can see that there is a trade-off between the success probability and the maximum correct probability, so the desired value of $p$ depends on the practical consideration. We can see from Fig. 2 that although the maximum correct probabilities for different nonorthogonal states are different, it is obvious that our scheme increases the maximum correct probability of two nonorthogonal states after passing the noise channel significantly compared with that without the FFC. And the larger $\theta$ is, the superior our scheme is.
It is noted that there is a kink in Fig. 2 (and also in the following figures in this section) which is due to the absolute values in Eq. (\ref{PC12}).
\begin{widetext}
\begin{center}
\includegraphics[width=15cm]{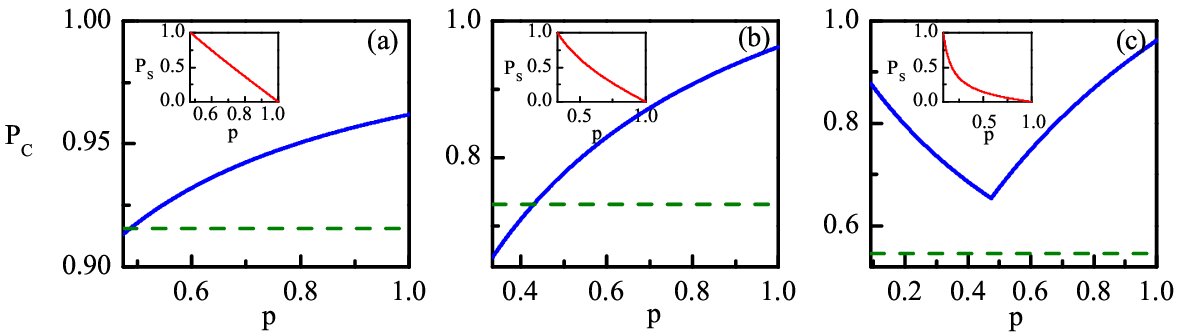}
\parbox{15cm}{\small{FIG. 3.} (Color online) The maximum correct probabilities $P_{C}^{fin}$ (Eq. (\ref{PCFIN=1})) (solid-blue lines) and $P_{C}^{mix}$ (Eq. (\ref{PC})) (dashed-olive lines) as functions of the measurement strength $p$  for (a) $r=0.1$; (b) $r=0.5$ and (c) $r=0.9$. The insets plot the corresponding success probabilities $P_{S}$ for our scheme. $\theta=6\pi/16,~ q_{+}=q_{-}=1/2$.}
\end{center}
\end{widetext}
\begin{center}
\includegraphics[width=8 cm]{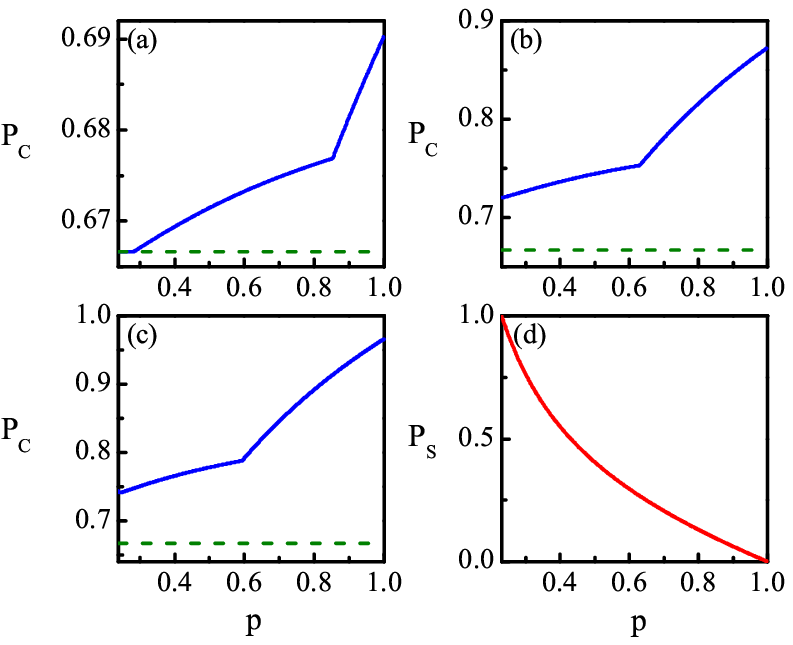}
\parbox{8cm}{\small{FIG. 4.} (Color online) The maximum correct probabilities $P_{C}^{fin}$ (Eq. (\ref{PCFIN})) (solid-blue lines) and $P_{C}^{mix}$ (Eq. (\ref{PCCC})) (dashed-olive lines) as functions of the measurement strength $p$  for (a) $\theta=\pi/16$; (b) $\theta=4\pi/16$ and (c) $\theta=6\pi/16$. Panel (d) plots the success probability $P_{S}$ corresponding to panels (a--c) for our scheme which is state independent. $r=0.7, ~q_{+}=1/3,~ q_{-}=2/3$.}
\end{center}

Besides, we also consider the effect of different degrees of amplitude damping on the maximum correct probability for our scheme and compare our result with that of the ME discrimination without the FFC. We find that our scheme can be always superior over that without the FFC for any degree of amplitude damping. As an example, we give our maximum correct probabilities for different degrees of amplitude damping, with a fixed pair of initial nonorthogonal states $\theta=6\pi/16$ and equal a prior probabilities $q_{+}=q_{-}=1/2$ in Fig. 3. It can be seen that our scheme greatly increases the maximum correct probability of two nonorthogonal states after passing the noise channel compared with that without the FFC for any degree of amplitude damping, even for heavy amplitude damping noise (e.g., $r=0.9$).

The above discussions of discrimination effect for our scheme are based on equal a prior probabilities. Now we consider unequal a prior probabilities. Similarly, we first plot the maximum correct probability and the success probability for our scheme as functions of the measurement strength $p$ in Fig. 4 for different pairs of nonorthogonal states, with a fixed degree of amplitude damping $r=0.7$ and unequal a prior probabilities $q_{+}=1/3,~ q_{-}=2/3$.
Compared with the ME discrimination without the FFC, our scheme can still reach a larger maximum correct probability for any two nonorthogonal states. And the larger $\theta$ is, the more superior our scheme is.

Next, we compare our scheme with that without the FFC for different degrees of amplitude damping, with a fixed pair of nonorthogonal states $\theta=6\pi/16$ and unequal a prior probabilities $q_{+}=1/3$,~ $q_{-}=2/3$ in Fig. 5. We can see that in the case of unequal a prior probabilities, our scheme still has an advantage for any degree of amplitude damping, even for heavy amplitude damping $r=0.9$. This feature of our scheme reveals its significance in quantum communication, for we can discriminate what the initial state sent is with a fairly large maximum correct probability even for heavy AD channel.

So the contribution of this discrimination scheme lies in the fact that, by choosing a proper measurement strength, the maximum correct probability of two nonorthogonal states after passing the AD channel can be highly increased compared with that without the FFC for any initial nonorthogonal states, any degree of amplitude damping and any prior probabilities. In particular, the larger $\theta$ is, the more advantageous our scheme is.

\begin{widetext}
\begin{center}
\includegraphics[width=15cm]{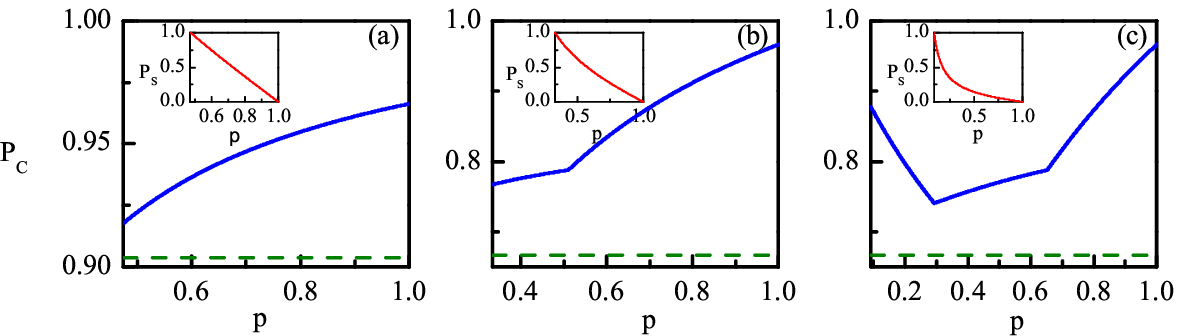}
\parbox{15cm}{\small{FIG. 5.} (Color online) The maximum correct probabilities $P_{C}^{fin}$ (Eq. (\ref{PCFIN})) (solid-blue lines) and $P_{C}^{mix}$ (Eq. (\ref{PCCC})) (dashed-olive lines) as functions of the measurement strength $p$  for (a) $r=0.1$; (b) $r=0.5$ and (c) $r=0.9$. The insets plot the corresponding success probabilities $P_{S}$ for our scheme. $\theta=6\pi/16, ~q_{+}=1/3, ~q_{-}=2/3$.}\label{fig5}
\end{center}
\end{widetext}
\section{optimal discrimination}
In Sec. \uppercase\expandafter{\romannumeral3}, we discuss the case of discrimination based on state protection. This case demands $p_{1}$ and $p_{2}$ having a specific relationship with $p$ for the purpose of protecting the initial nonorthogonal pure states, and based on which the two protected nonorthogonal states after passing the noise channel can be further discriminated. However, this discrimination strategy is not an optimal one.
\begin{widetext}
\begin{center}
\includegraphics[width=15cm]{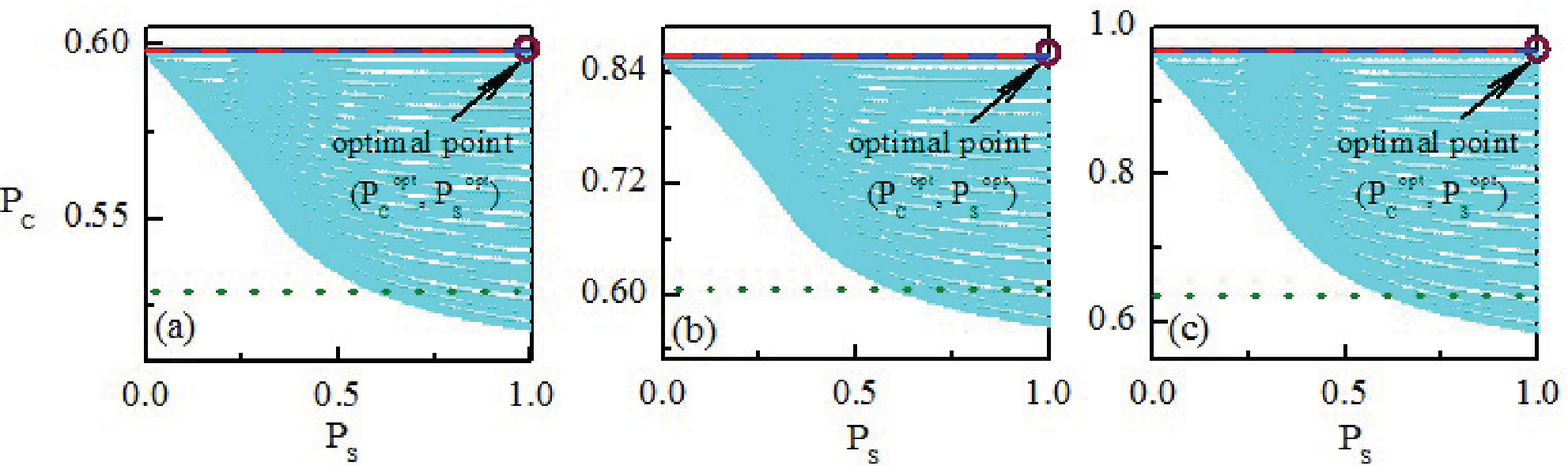}
\parbox{15cm}{\small{FIG. 6.} (Color online) $P_{C}^{fin}-P_{S}$ diagrams for our scheme (blue region) with $P_{C}^{mix}$ (Eq. (\ref{PC})) (dotted-olive lines) and $P_{C}^{pure}$ (Eq. (\ref{pcpure=})) (dashed-red lines) for (a) $\theta=\pi/16$; (b) $\theta=4\pi/16$ and (c) $\theta=6\pi/16$. $r=0.7,~ q_{+}=q_{-}=1/2$. The blue lines represent our optimal lines which coincide with $P_{C}^{pure}$ and the purple circles represent our optimal points.}\label{fig6}
\end{center}
\end{widetext}

Now in this section, the only purpose for us is to discriminate between two nonorthogonal states after passing the noise channel with a large maximum correct probability as much as possible. Actually all the measurement strengths $p,~ p_{1},~ p_{2}$ can be varied independently from 0 to 1, and each group of the measurement strengths ($p,~ p_{1},~ p_{2}$) uniquely determines its corresponding maximum correct probability and success probability according to Eqs. (\ref{PCFIN}), (\ref{PS}) denoted by ($P_{C}^{fin},~P_{S}$). It is conceivable that all ($P_{C}^{fin},~P_{S}$) would constitute $P_{C}^{fin}$--$P_{S}$ diagram. We first show the $P_{C}^{fin}$--$P_{S}$ diagram of several nonorthogonal states, with a fixed degree of amplitude damping $r=0.7$ and equal a prior probabilities $q_{+}=q_{-}=1/2$ in Fig. 6. It is noted that for a given succuss probability, the point ($P_{C}^{fin},~P_{S}$), at which the value of $P_{C}^{fin}$ is the largest, is distributed on the boundary of the diagram denoted by a blue line. We call this boundary line as optimal line of discrimination. Here we should emphasize that different groups of measurement strengths ($p,~p_{1},~p_{2}$) determine the different final states $\rho^{fin(i)}$ (Eq. (\ref{rhofini})), hence corresponding to different maximum correct probabilities $P_{C}^{fin}$ (Eq. (\ref{PCFIN})). That is, different groups of measurement strengths ($p,~p_{1},~p_{2}$) correspond to different values of $P_{C}^{fin}$. From Fig. 6 we can see that on the optimal line although the success probability varies with the varying $p,~ p_{1}$, and $p_{2}$, the maximum correct probability remains unchanged, and is equal to that of the two initial nonorthogonal pure states. In this case, we define the optimal discrimination as the one with both the largest maximum correct probability and the largest success probability, corresponding to an optimal point of discrimination ($P_{C}^{opt}, ~P_{S}^{opt}$), which is denoted by a purple circle in Fig. 6 (later we will analytically derive the condition of optimal discrimination, i.e., the optimal group of measurement strengths ($p,~ p_{1},~ p_{2}$)). In addition, we also compare our scheme (optimal line of discrimination) with that without the FFC, we find that the maximum correct probability in our scheme is much larger than that of the ME discrimination without the FFC. And the larger $\theta$ is, the more advantage our scheme has. For example, when $\theta=6\pi/16$, the increased percentage of the maximum correct probability of our scheme (optimal line of discrimination) compared with that without the FFC can arrive at more than 50\%.

Besides, we also consider the effect of different degrees of amplitude damping on the maximum correct probability for our scheme and numerically compare our scheme with that without the FFC. In Fig. 7 (and also in the following figures), we just plot the optimal line of $P_{C}^{fin}$-$P_{S}$ diagram, which is clear enough to illustrate the effect of the optimal discrimination. As an example, we plot the optimal line for different degrees of amplitude damping, with a fixed pair of initial nonorthogonal states $\theta=6\pi/16$, and equal a prior probabilities $q_{+}=q_{-}=1/2$ in Fig. 7. We find that the maximum correct probability in our scheme is independent of the degree of amplitude damping and we can also see that our scheme has a greater advantage over that without the FFC when the amplitude damping is heavier. Here it should be noted that the maximum correct probability of two nonorthogonal states after passing the noise channel for any degree of amplitude damping by our scheme can achieve that of the two initial nonorthogonal pure states.
\begin{center}
\includegraphics[width=8 cm]{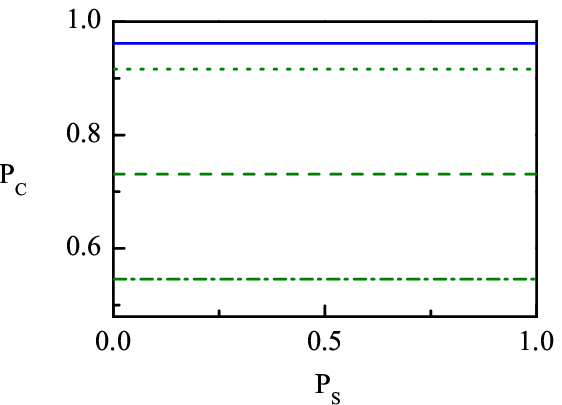}
\parbox{8cm}{\small{FIG. 7.} (Color online) $P_{C}^{fin}-P_{S}$ diagram for different degrees of amplitude damping. The solid-blue line represents both our scheme for $r=0.1$, $r=0.5$, $r=0.9$ and $P_{C}^{pure}$ (Eq. (\ref{pcpure=})) which coincide; $P_{C}^{mix}$ (Eq. (\ref{PC})) is represented by dotted-olive line ($r=0.1$), dashed-olive line ($r=0.5$) and dashed dotted-olive line ($r=0.9$). $\theta=6\pi/16,~ q_{+}=q_{-}=1/2$. }
\end{center}
\begin{widetext}
\begin{center}
\includegraphics[width=15cm]{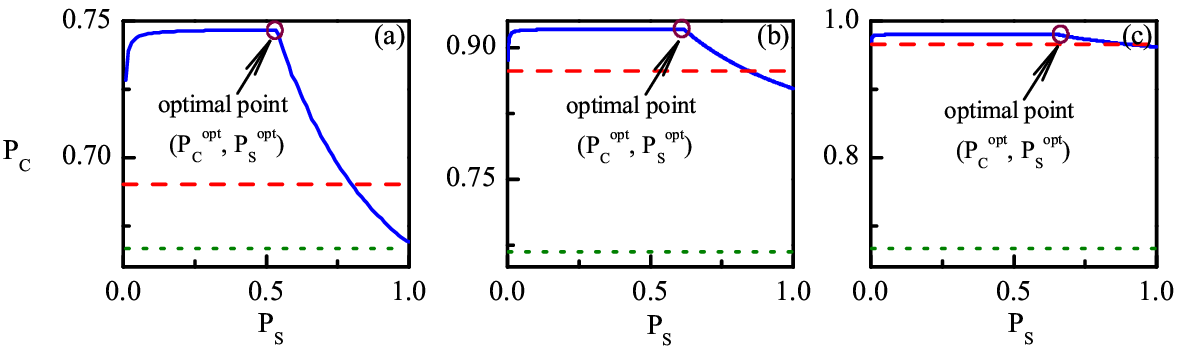}
\parbox{15cm}{\small{FIG. 8.} (Color online) $P_{C}^{fin}-P_{S}$ diagrams for our scheme (solid-blue lines) with $P_{C}^{mix}$ (Eq. (\ref{PCCC})) (dotted-olive lines) and $P_{C}^{pure}$ (Eq. (\ref{pcpure})) (dashed-red lines) for (a) $\theta=\pi/16$; (b) $\theta=4\pi/16$ and (c) $\theta=6\pi/16$. $r=0.7,~ q_{+}=1/3, ~q_{-}=2/3$. The purple circles represent our optimal points.}\label{fig8}
\end{center}
\end{widetext}

Similar to the discussion in Sec. \uppercase\expandafter{\romannumeral3}, we then consider the case of unequal a prior probabilities. First, we plot the optimal line of $P_{C}^{fin}$-$P_{S}$ diagram for different pairs of nonorthogonal states, with a fixed degree of amplitude damping $r=0.7$ and unequal a prior probabilities $q_{+}=1/3$, $q_{-}=2/3$ in Fig. 8 as an example. We can see from Fig. 8 that with the measurement strengths $p,~p_{1}$, and $p_{2}$ varying, the maximum correct probability on the optimal line first remains unchanged and then decreases, but is still larger than that of the ME discrimination without the FFC. It can be seen that there exists a trade-off between the maximum correct probability and the success probability. Our aim is to first increase the maximum correct probability and then pursue the success probability as large as possible. Based on this, we define the optimal discrimination as the one which first ensures the largest value of $P_{C}^{fin}$ and then pursues the success probability as large as possible. This optimal discrimination corresponds to the optimal point of discrimination ($P_{C}^{opt}, ~P_{S}^{opt}$) denoted by a purple circle in Fig. 8 (later we will give the condition of the optimal discrimination, i.e., the optimal group of measurement strengths ($p,~ p_{1},~ p_{2}$)). The remarkable fact is that the maximum correct probability of two nonorthogonal states after passing the noise channel in our scheme can be even larger than that of the two initial nonorthogonal pure states in a probabilistically way for unequal a prior probabilities. Compared with that without the FFC, our scheme still has an obvious advantage. And we can also see that the maximum correct probability in our scheme increases as $\theta$ increases for each fixed success probability,
so the larger $\theta$ is, the more advantage our scheme has.
\begin{center}
\includegraphics[width=8 cm]{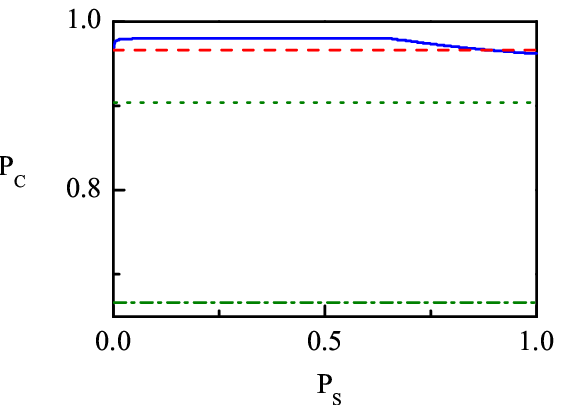}
\parbox{8cm}{\small{FIG. 9.} (Color online) $P_{C}^{fin}-P_{S}$ diagram for different degrees of amplitude damping. The solid-blue line represents our scheme for $r=0.1$, $r=0.5$ and $r=0.9$ which coincide; The dashed-red line represents $P_{C}^{pure}$ (Eq. (\ref{pcpure})); $P_{C}^{mix}$ (Eq. (\ref{PCCC})) is represented by dotted-olive line ($r=0.1$), dashed dotted-olive line ($r=0.5$ and $r=0.9$ which coincide). $\theta=6\pi/16, ~ q_{+}=1/3, ~q_{-}=2/3$.}
\end{center}

And then we consider the effect of different degrees of amplitude damping on the maximum correct probability. In Fig. 9, we plot the optimal line of $P_{C}^{fin}$-$P_{S}$ diagram for different degrees of amplitude damping, with a fixed pair of initial nonorthogonal states $\theta=6\pi/16$ and unequal a probabilities $q_{+}=1/3$, $q_{-}=2/3$ as an example. We can see that the maximum correct probability in our scheme is independent of the degree of amplitude damping and compared with that without the FFC, our scheme still has an advantage for different degrees of amplitude damping in the case of unequal a prior probabilities. Moreover, the maximum correct probability of two nonorthogonal states after passing the noise channel by our scheme can be larger than that of the two initial nonorthogonal pure states for any degree of amplitude damping.

Now we give the condition of the optimal discrimination, i.e., the optimal group of measurement strengths ($p,~p_{1},~p_{2}$). First we consider the case of equal a prior probabilities, $q_{+}$=$q_{-}$=1/2.
We begin with the expression of $P_{C}^{fin}$ (Eq. (\ref{PCFIN=})) and discuss it in three circumstances:

(a) When $r>\frac{p}{1-p}$,
\begin{equation}\label{pc1}
    P_{C}^{fin}=\frac{1}{2}+\frac{2-4p+(p_{1}+p_{2})(p-(1-p)r)}{2\sin^{-1}\theta(2-(p_{1}+p_{2})(p(1-r)+r))},
\end{equation}
\begin{equation}\label{daopc1}
    \frac{\partial P_{C}^{fin}}{\partial p}=\frac{\sin\theta(2-p_{1}-p_{2})((p_{1}+p_{2})r-2)}{(2-(p_{1}+p_{2})(p(1-r)+r))^{2}}.
\end{equation}
From Eq. (\ref{daopc1}), we can see that if

(a-1) $p_{1}$=$p_{2}$=1,  $\partial P_{C}^{fin}/\partial p$=0. Substituting $p_{1}$=$p_{2}$=1 into Eq. (\ref{pc1}), we obtain that
\begin{equation}\label{pc11}
    P_{C}^{fin}=\frac{1}{2}(1+\sin\theta)=P_{C}^{pure},
\end{equation}
which equals to the maximum correct probability of discriminating two initial nonorthogonal pure states $P_{C}^{pure}$ (Eq. (\ref{pcpure=})), and the corresponding success probability of our scheme (Eq. (\ref{PS=})) is
\begin{equation}\label{ps11}
    P_{S}=(1-p)(1-r).
\end{equation}

(a-2) $p_{1},~ p_{2}\neq1$, $\partial P_{C}^{fin}/\partial p<0$ which means that $P_{C}^{fin}$ is monotonously decreasing with $p$, when we choose $p$=0,
\begin{equation}\label{pc12}
    P_{C}^{fin}=\frac{1}{2}(1+\sin\theta)=P_{C}^{pure},
\end{equation}
and the corresponding success probability is
\begin{equation}\label{ps12}
    P_{S}=1-\frac{1}{2}r(p_{1}+p_{2}).
\end{equation}

(b) When $r<\frac{p}{1-p}$,
\begin{equation}\label{pc2}
    P_{C}^{fin}=\frac{1}{2}+\frac{2+4r(p-1)+(p_{1}+p_{2})(r-p(r+1))}{2\sin^{-1}\theta(2-(p_{1}+p_{2})(p(1-r)+r))},
\end{equation}
and
\begin{equation}\label{daopc2}
    \frac{\partial P_{C}^{fin}}{\partial p}=\frac{\sin\theta(2-p_{1}-p_{2})^{2}r}{(2-(p_{1}+p_{2})(p(1-r)+r))^{2}}.
\end{equation}
From Eq. (\ref{daopc2}), we can see that if

(b-1) $p_{1}$=$p_{2}$=1,  $\partial P_{C}^{fin}/\partial p$=0 which is the same as case (a-1).

(b-2) $p_{1},~ p_{2}\neq1$, $\partial P_{C}^{fin}/\partial p>0$ which means that $P_{C}^{fin}$ is monotonously increasing with $p$, and when we choose $p$=1,
\begin{equation}\label{pc22}
    P_{C}^{fin}=\frac{1}{2}(1+\sin\theta)=P_{C}^{pure},
\end{equation}
and the corresponding success probability is
\begin{equation}\label{ps22}
    P_{S}=1-\frac{1}{2}(p_{1}+p_{2}).
\end{equation}

(c) When $r=\frac{p}{1-p}$,
\begin{equation}\label{pc3}
    P_{C}^{fin}=\frac{1}{2}+\frac{2p-1}{2\sin^{-1}\theta (p(p_{1}+p_{2})-1)},
\end{equation}
\begin{equation}\label{daopc3}
     \frac{\partial P_{C}^{fin}}{\partial p_{1}}=\frac{\partial P_{C}^{fin}}{\partial p_{2}}=\frac{p(1-2p)}{(p(p_{1}+p_{2})-1)^{2}},
\end{equation}
where $p<1/2$ due to $r=\frac{p}{1-p}<1$. So from Eq. (\ref{daopc3}), we can see that both $\partial P_{C}^{fin}/\partial p_{1}$ and $\partial P_{C}^{fin}/\partial p_{2}$ are greater than zero, i.e., $P_{C}^{fin}$ is monotonously increasing with $p_{1}$ and $p_{2}$. When we choose
$p_{1}$=$p_{2}$=1, it is the same as case (a-1).

In conclusion, in all the cases, the maximum correct probability can be equal to that of the two initial nonorthogonal pure states.
Moreover the success probability in cases (a-2) and (b-2) can arrive at 1 when we choose $p_{1}$=$p_{2}$=0. So the condition of the optimal discrimination, i.e., the optimal group of measurement strengths, can be either ($p=0,~p_{1}=0,~p_{2}=0$) or ($p=1,~p_{1}=0,~p_{2}=0$), which corresponds to the optimal point of discrimination (see Fig. 6) with the maximum correct probability $P_{C}^{opt}=P_{C}^{pure}$ and the success probability $P_{S}^{opt}=1$.

Furthermore, in the case of unequal a prior probabilities, we numerically find the condition of the optimal discrimination, i.e., the optimal group of measurement strengths $(p,~p_{1},~p_{2})$ which corresponds to the optimal point of discrimination (see Fig. 8), with respect to different cases of prior probabilities:
\begin{equation}\label{optimal}
\begin{split}
   q_{+}>q_{-}, ~  (p=1,~ p_{1}=0,~ p_{2}=1);\\
   q_{+}<q_{-},  ~   (p=1,~ p_{1}=1,~ p_{2}=0),\\
 \end{split}
\end{equation}
which are independent of the initial nonorthogonal states and the degrees of amplitude damping, only determined by the prior probabilities.

And under these optimal groups of measurement strengths ($p,~p_{1},~p_{2}$) (Eq. (\ref{optimal})), we can obtain analytical expressions of the maximum correct probability $P_{C}^{opt}$ and the corresponding success probability $P_{S}^{opt}$ of the optimal point for any pair of nonorthogonal states according to Eqs. (\ref{PCFIN}), (\ref{PS}):
\begin{equation}\label{PCOPT}
     P_{C}^{opt}=\frac{1}{2}(1+\frac{|\chi|+\sin\theta}{1+\sin\theta|\chi|}),
\end{equation}
where $\chi$ has been defined in Eq. (\ref{PCCC}),
and
\begin{equation}\label{PSOPT}
P_{S}^{opt}=\frac{1}{2}(1+\sin\theta|\chi|).
\end{equation}
It should be noted that Eq. (\ref{PCOPT}) is irrelevant to the degree of amplitude damping, so generally the heavier the amplitude damping is, the more advantageous our scheme is.

Besides, we numerically compare the analytical expression of $P_{C}^{opt}$ (Eq. (\ref{PCOPT})) with that of two initial nonorthogonal pure states, $P_{C}^{pure}=1/2(1+\sqrt{1-4q_{+}q_{-}\cos^2\theta})$ (Eq. (\ref
{pcpure})), and find
\begin{equation}\label{dayu}
    P_{C}^{opt}>P_{C}^{pure}
\end{equation}
for any pair of initial nonorthogonal states and any degree
of amplitude damping.

So the significance of this optimal discrimination lies in an remarkable discrimination effect in the presence of the environmental noise, that is, by using this discrimination scheme, one can make the maximum correct probability equal to that of the two initial nonorthogonal pure states in the case of equal a prior probabilities, and even larger than that of the two initial nonorthogonal pure states in probabilistically way in the case of unequal a prior probabilities, while this cannot be achieved by the ME discrimination without the FFC.

Finally, in order to comprehensively consider the two figures of merit, i.e., the maximum correct probability $P_{C}^{fin}$ and the success probability $P_{S}$ in our scheme, we define a new quantity which equals to the product of them \cite{Qiao2013}, i.e., $P_{SC}=P_{S}P_{C}^{fin}=P_{S_{1}}P_{C}^{fin(1)}+P_{S_{2}}P_{C}^{fin(2)}$. And we find that in the case of the discrimination based on state protection, for equal a prior probabilities, $P_{SC}$ can be larger than $P_{C}^{mix}$ in the regime $P_{S}\rightarrow1$ only for heavier amplitude damping $r$ (see Fig. 2 and Fig. 3) and for unequal a prior probabilities or in the case of optimal discrimination, $P_{SC}$ can be larger than $P_{C}^{mix}$ also in the regime $P_{S}\rightarrow1$ for any degree of amplitude damping $r$ (see Fig. 5, Fig. 7 and Fig. 9).

\section{Conclusions}
In this paper, we consider how to improve the effect of discrimination of two nonorthogonal states
after passing the AD channel. To this end, we propose
a scheme by using the FFC to realize a better effect of discrimination for any pair of nonorthogonal states and
any degree of amplitude damping. The core idea of our scheme is to make use of the pre-weak measurement and  unitary operation to make the initial nonorthogonal states into the ones almost immune to the AD channel, and then to restore the initial states sent originally after their transmission by means of the reversed  unitary operation and reversed post-weak measurement. Based on this we further discriminate between the two transmitted nonorthogonal states by using the ME discrimination. Concretely speaking, we consider two cases: (i) discrimination based on state protection and (ii) optimal discrimination. For the former, the purpose is to protect the initial nonorthogonal pure states, and further discriminate between the two protected nonorthogonal states after passing the noise channel. Based on this, the discrimination effect of the two transmitted nonorthogonal mixed states can be close to that of the two initial nonorthogonal pure states especially when $p\rightarrow1$, but at the cost of the substantial decrease of the success probability; For the latter, the only purpose is to discriminate between the two transmitted nonorthogonal mixed states with a large maximum correct probability as much as possible. In this case, when $q_{+}=q_{-}=1/2$, the maximum correct probability can be equal to that of the two initial nonorthogonal pure states in a deterministic way; when $q_{+}\neq q_{-}$, the maximum correct probability can be even larger than that of the two initial nonorthogonal pure states in a probabilistically way. But for both cases the discrimination effect can be highly improved compared with that without the FFC.

So the contribution of this paper to the state discrimination field mainly lies in the following two aspects: the first is that we take the environmental noise into account in the state transmission process, which is closer to the reality in quantum communication. And based on this consideration, we propose a specific scheme by utilizing the FFC to counteract the detrimental impact of the environmental noise on the discrimination effect hence to increase the maximum correct probability; the second is that, by using our optimal discrimination scheme, one can completely counteract the impact of the environmental noise on the effect of discrimination of two nonorthogonal states in the case of equal a prior probabilities or even improve the effect of discrimination of two initial nonorthogonal pure states in the case of unequal a prior probabilities, while this cannot be achieved by the ME discrimination without the FFC. Besides, our scheme can be advantageous over the ME discrimination without the FFC for any two initial nonorthogonal states, any degree of amplitude damping noise and any prior probabilities. Moreover, the heavier the amplitude damping is, the more superior our scheme is.

\begin{acknowledgments}
This work was supported by the National Natural Science
Foundation of China (Grants No. 11274043, 11375025) and the National Science Foundation of
Shandong Province, China (Grants No. ZR2011FL009).

\end{acknowledgments}

\bibliography{SFDM5BIBR2}

\end{document}